\documentclass[a4paper,twocolumn,nofootinbib,floatfix,secnumarabic]{revtex4}
\usepackage{graphicx}

\begin{document}

\title{Collaborative Tagging and Semiotic Dynamics}

\author{Ciro Cattuto}
\email{ciro.cattuto@roma1.infn.it}
\affiliation{Museo Storico della Fisica e Centro Studi e Ricerche ``Enrico Fermi''\\
Compendio Viminale, 00184 Roma, Italy}
\affiliation{Dipartimento di Fisica, Universit\`{a} di Roma ``La Sapienza''\\
P.le A. Moro, 2, 00185 Roma, Italy}
\author{Vittorio Loreto}
\author{Luciano Pietronero}
\affiliation{Dipartimento di Fisica, Universit\`{a} di Roma ``La Sapienza''\\
P.le A. Moro, 2, 00185 Roma, Italy}

\date{\today}

\begin{abstract}
Collaborative tagging has been quickly gaining ground because of its ability to recruit
the activity of web users into effectively organizing and sharing vast amounts of information.
Here we collect data from a popular system and investigate
the statistical properties of tag co-occurrence. 
We introduce a stochastic model of user behavior embodying two main aspects
of collaborative tagging: (i) a frequency-bias mechanism related to the idea
that users are exposed to each other's tagging activity;
(ii) a notion of memory -- or aging of resources --
in the form of a heavy-tailed access to the past state of the system.
Remarkably, our simple modeling is able to account quantitatively
for the observed experimental features, with a surprisingly high accuracy.
This points in the direction of a universal behavior of users,
who -- despite the complexity of their own cognitive processes and
the uncoordinated and selfish nature of their tagging activity --
appear to follow simple activity patterns.
\end{abstract}

\maketitle

\begin{figure}[htb]
\includegraphics[width=\columnwidth,keepaspectratio]{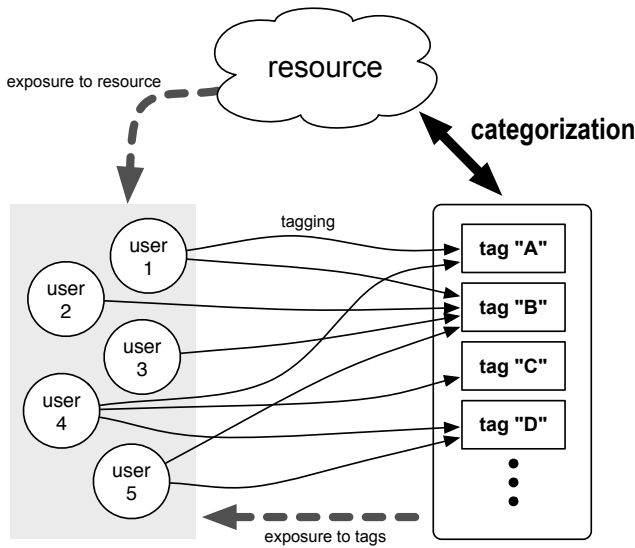}
\caption{
Schematic depiction of the collaborative tagging process: web users are exposed to a resource and freely associate tags with it. Their interaction with the system also exposes them to tags previously entered by themselves and by other users. The aggregated activity of users leads to an emergent categorization of resources in terms of tags shared by a community.
\label{fig1}
}
\end{figure}

\section{Introduction}
Recently, a new paradigm has been quickly gaining ground on the World-Wide Web:
Collaborative Tagging~\cite{b1,b2,b4}. In web applications like
\textit{del.icio.us}\footnote{\url{http://del.icio.us}},
\textit{Flickr}\footnote{\url{http://www.flickr.com}},
\textit{CiteULike}\footnote{\url{http://www.citeulike.org}},
\textit{Connotea}\footnote{\url{http://www.connotea.org}},
users manage, share and browse collections of online resources by enriching
them with semantically meaningful information in the form of freely
chosen text labels (\textit{tags}). The paradigm of collaborative tagging
has been successfully deployed in web applications designed to organize
and share diverse online resources such as bookmarks,
digital photographs, academic papers, music and more.
Web users interact with a collaborative tagging
system by posting content (\textit{resources}) into the system,
and associating text strings (\textit{tags}) with that content,
as shown in Fig.~\ref{fig1}.
At the global level the set of tags, though determined with no explicit coordination,
evolves in time and leads towards patterns of terminology usage
that are shared by the entire user community. Hence one observes
the emergence of a loose categorization system
-- commonly referred to as \textit{folksonomy} -- that can be effectively used
to navigate through a large and heterogeneous body of resources.

Focusing on tags as basic dynamical entities, the process of collaborative
tagging falls within the scope of Semiotic Dynamics~\cite{b5,b6},
a new field that studies how populations of humans or agents can establish
and share semiotic systems, typically driven by their use in communication.
Indeed the emergence of a folksonomy exhibits dynamical aspects
also observed in human languages~\cite{b15,b16}, such as the crystallization
of naming conventions, competition between terms, takeovers by neologisms, and more.

In the following we adopt the point of view of complex systems science
and try to understand how the ``microscopic'' tagging activity of users
causes the emergence of the high-level features we observe for the ensuing folksonomy.
We ground our analysis on actual tagging data extracted from \textit{del.icio.us}
and \textit{Connotea}
and use standard statistical tools to gain insights into the underlying tagging dynamics.
Based on this, we introduce a simple stochastic model for the tagging behavior
of an ``average'' user, and show that such a model -- despite its simplicity --
is able to reproduce extremely well some of the observed properties.
We close giving an interpretation of the model parameters
and pointing out directions for future research.

\section{Experimental Data}
The activity of users interacting with a collaborative tagging system
consists of either navigating the existing body of resources by using
tags, or adding new resources to the system. In order to add a new
resource to the system, the user is prompted for a reference to the
resource and a set of tags to associate with it. Thus the basic unit
of information in a collaborative tagging system is a {\tt(user,
resource, \{tags\})} triple, here referred to as \textit{post}. Tagging events
build a tri-partite graph (with partitions corresponding to
users, resources and tags, respectively) and such a graph can be subsequently
used as a navigation aid in browsing tagged information.
Usually a post contains also a temporal marker indicating the physical time
of the tagging event, so that temporal ordering can be preserved
in storing and retrieving posts.

Our analysis will focus on \textit{Del.icio.us}, for several reasons:
i) it was the very first system to deploy the ideas and technologies
of collaborative tagging, and the paradigmatic character it acquired
makes it a natural starting point for any quantitative study.
ii) because of its popularity, it has a large community of active users
and comprises a precious body of raw data on the static and dynamical
properties of a folksonomy.
iii) it is a \textit{broad folksonomy}~\cite{vanderwal},
and single tagging events (posts) retain their identity
and can be individually retrieved.
This affords unimpeded access to the ``microscopic'' dynamics
of collaborative tagging, providing the opportunity to make contact
between emergent behaviors and low-level dynamics.
It also allows to define and measure the multiplicity (or frequency)
of tags in the context of a single resource. Contrary to this,
popular sites falling in the \textit{narrow folksonomy} class
(\textit{Flickr}, for example) foster a different model of user interaction,
where tags are mostly applied by the content creator, no notion of tag multiplicity
is available in the context of a resource, and no access is given
to the raw sequence of tagging events.

\begin{figure*}[htb]
\hspace{-1.0cm}
\includegraphics[width=\textwidth,keepaspectratio]{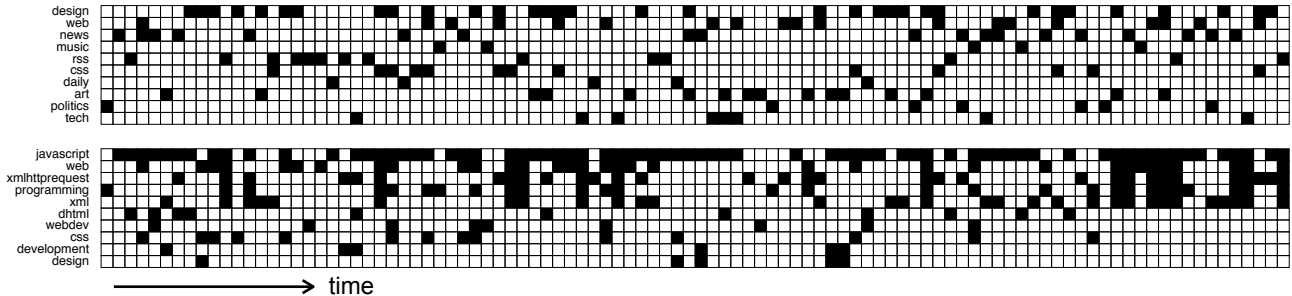}
\caption{
Tagging activity: a time-ordered sequence of tagging events is graphically rendered by marking
the tags co-occurring with \textit{blog} (top panel) or \textit{ajax} (bottom panel)
in an experimental sequence of posts on \textit{del.icio.us}.
In each panel, columns represent single tagging events (posts)
and rows correspond to the $10$ most frequent tags
co-occurring with either \textit{blog} (top panel) or \textit{ajax} (bottom panel).
$100$ tagging events are shown in each panel, temporally ordered from left to right.
Only posts involving at least one of the $10$ top-ranked tags are shown.
For each tagging event (column), a filled cell marks the presence of the tag in the corresponding row,
while an empty cell indicates its absence. A qualitative difference between \textit{blog} (top panel)
and \textit{ajax} (bottom panel) is clearly visible, where a higher density at low-rank tags
characterizes the semantically narrower \textit{ajax} term. This corresponds to the steeper low-rank
behavior observed in the frequency-rank plot for \textit{ajax} (Fig.~\ref{fig2b}).
\label{fig2a}
}
\end{figure*}

On studying \textit{Del.icio.us} we adopt a tag-centric view of
the system, that is we investigate the evolving relationship between a
given tag and the set of tags that co-occur with it. In line with
our focus on semiotic dynamics, we factor out the detailed identity of
the users involved in the process, and only deal with streams
of tagging events and their statistical properties. To perform automated
data collection of raw data we use a custom web (HTTP) client that
connects to \textit{del.icio.us} and navigates the system's interface
as an ordinary user would do, extracting the relevant metadata and
storing it for further post-processing. \textit{del.icio.us} allows
the user to browse its content by tag: our client requests the
web page associated with the tag under study and uses an HTML
parser to extract the post information (user, resource, tags, time stamp)
from the returned HTML code. Fig.~\ref{fig2a} graphically depicts
the raw data we gather, for the case of two popular tags on
\textit{del.icio.us}.
Table~\ref{table1} describes the datasets we used for the present analysis.

\begin{figure*}[htb]
\vspace*{0.5cm}
\hspace*{-1.75cm}
\includegraphics[width=0.9\textwidth,keepaspectratio]{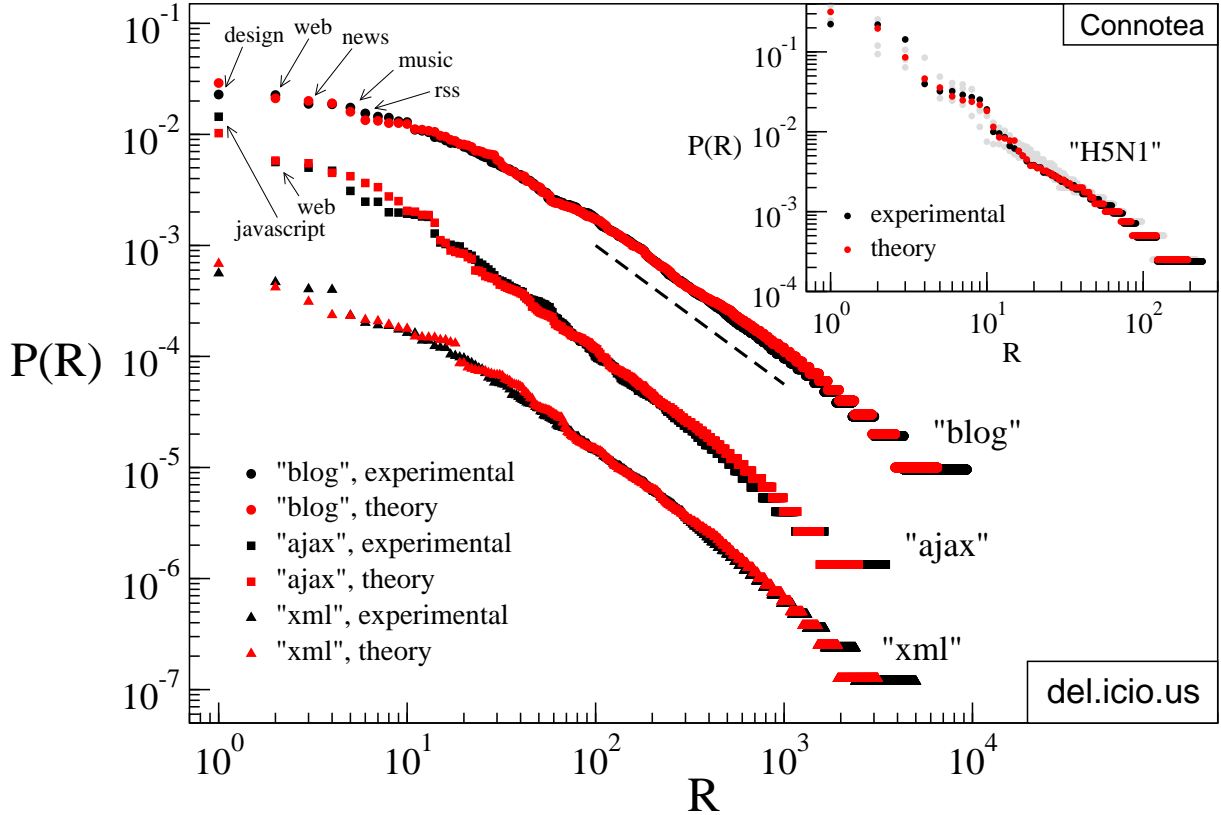}
\caption{
Frequency-rank plots for tags co-occurring with a selected tag: experimental data (black symbols)
are shown for \textit{del.icio.us} (circles for tags co-occurring with the popular tag \textit{blog},
squares for \textit{ajax} and triangles for \textit{xml}) and \textit{Connotea}
(inset, black circles for the \textit{H5N1} tag). For the sake of clarity, the curves for \textit{ajax}
and \textit{xml} are shifted down by one and two decades, respectively.
Details about the experimental datasets are reported in Table~\ref{table1}.
All curves exhibit a power-law decay for high ranks
(a dashed line corresponding to the power law $R^{-5/4}$ is provided as an aid for eye)
and a shallower behavior for low ranks.
To make contact with Fig.~\ref{fig2a},
some of the highest-frequency tags co-occurring with \textit{blog} and \textit{ajax} are explicitly indicated
with arrows. Red symbols are theoretical data obtained by computer simulation of the stochastic process
described in the text (Fig.~\ref{fig3}). The parameters of the model, i.e. the probability $p$,
the
memory parameter $\tau$ and the initial number of words $n_0$ were adjusted
to match the experimental data, giving approximately $p=0.06$, $\tau=100$  and $n_0=100$ for \textit{blog},
$p=0.03$, $\tau=20$ and $n_0=50$ for \textit{ajax}, and $p=0.034$, $\tau=40$ and $n_0=110$ for \textit{xml}.
Inset: \textit{Connotea} is a much younger system than \textit{del.icio.us} and the corresponding dataset is smaller and noisier. Nevertheless, a good match with experimental data can be obtained
for $p=0.05$, $\tau=120$ and $n_0=7$ (red circles), demonstrating that our model also applies
to the early stages of development of a folksonomy.
Gray circles correspond to different realizations of the simulated dynamics.
\label{fig2b}
}
\end{figure*}

\begin{table}
\caption{\label{table1}
Statistics of the datasets used for the co-occurrence analysis. For each tag in the first column we report the number of posts marked with that tag, the number of total and distinct tags co-occurring with it,
and the corresponding number of resources.
The data were retrieved during May 2005.
\vspace{0.1cm}
}
\begin{tabular}{|l|l|l|l|l|}
\hline
Tag	& No. posts & No. tags & No. distinct tags & No. resources\\ \hline
Blog &	37974 & 124171 & 10617 &16990\\ \hline
Ajax & 33140 & 108181 &	4141 &	2995\\ \hline
Xml & 24249 & 108013 & 6035	& 7364\\ \hline
H5N1 & 981 & 5185 & 241 & 969 \\
\hline
\end{tabular}
\end{table}

\section{Data analysis}
Here we analyze data from \textit{del.icio.us} and \textit{Connotea}
and investigate the statistical properties of tag association.
Specifically, we select a semantic context by extracting
the resources
associated with a given tag $X$ and study the statistical distribution
of tags co-occurring with $X$ (see Table 1). Fig.~\ref{fig2a}
graphically illustrates the associations between tags and posts,
and Fig.~\ref{fig2b} reports the frequency-rank distributions
for the tags co-occurring with a few selected ones.
The high-rank tail of the experimental curves displays a power-law behavior,
signature of an emergent hierarchical structure,
corresponding to a generalized Zipf's law~\cite{b7}
with an exponent between $1$ and $2$.
Since power laws are the standard signature of self-organization
and of human activity~\cite{b8,b9}, the presence of a power-law tail is not surprising.
The observed value of the exponent, however, deserves further investigation
because the mechanisms usually invoked to explain Zipf's law and its generalizations~\cite{b12}
don't look very realistic for the case at hand, and a mechanism grounded on experimental
data should be sought.

Moreover, the low-rank part of the frequency-rank curves exhibits a flattening
typically not observed in systems strictly obeying Zipf's law.
Several aspects of the underlying complex dynamics may be responsible
for this feature: on the one hand this behavior points to the existence
of semantically equivalent and possibly competing high-frequency tags
(e.g. \textit{blog} and \textit{blogs}).
More importantly, this flattening behavior may be ascribed
to an underlying hierarchical organization of tags co-occurring
with the one we single out:
more general tags (semantically speaking)
will tend to co-occur with a larger number of other tags.
In this scenario, we expect a shallower behavior for tags co-occurring
with generic tags (e.g. \textit{blog})
and a steeper behavior for semantically narrow tags
(e.g. \textit{ajax}, see also Fig.~\ref{fig2a}).
To better probe the validity of this interpretation, we investigate
the co-occurrence relationship that links high-rank tags,
lying well within the power-law tail, with low-rank tags
located in the shallow part of the distribution.
Our observations (see section \ref{highlowrank})
point in the direction of a non-trivial
hierarchical organization emerging out of the collective tagging activity,
with each low-rank tag leading its own hierarchy of semantically related
higher-rank tags, and all such hierarchies merging into the overall
power-law tail.

\section{A Yule-Simon's model with long-term memory}
We now aim at gaining a deeper insight into the phenomenology reported above.
In order to model the observed frequency-rank behavior for the full range
of ranking values, we introduce a new version of the ``rich-get-richer''
Yule-Simon's stochastic model~\cite{b10,b11} by enhancing it with a fat-tailed
memory kernel. The original model can be described as the construction
of a text from scratch. At each discrete time step one word is appended to the text:
with probability $p$ the appended word is a new word,
never occurred before, while with probability $1-p$ one word
is copied from the existing text, choosing it with a probability proportional
to its current frequency of occurrence. This simple process yields
frequency-rank distributions that display a power-law tail with exponent $\alpha=1-p$,
lower than the exponents we observe in actual data.
This happens because the Yule-Simon process has no notion of ``aging'',
i.e. all positions within the text are regarded as identical.

In our construction we moved from the observation that
actual users are exposed in principle to all the tags stored in the system
(like in the original Yule-Simon model) but the way in which they choose
among them, when tagging a new resource,
is far from being uniform in time
(see also \cite{b17,mendes}). It seems more realistic to assume that users
tend to apply recently added tags more frequently than old ones,
according to a memory kernel which might be highly skewed.
Indeed, recent findings about human activities~\cite{b9} support the idea that
the access pattern to the past of the system should be fat-tailed,
suggesting a power-law memory kernel.

\begin{figure*}[htb]
\hspace*{-1.0cm}
\includegraphics[width=0.85\textwidth,keepaspectratio]{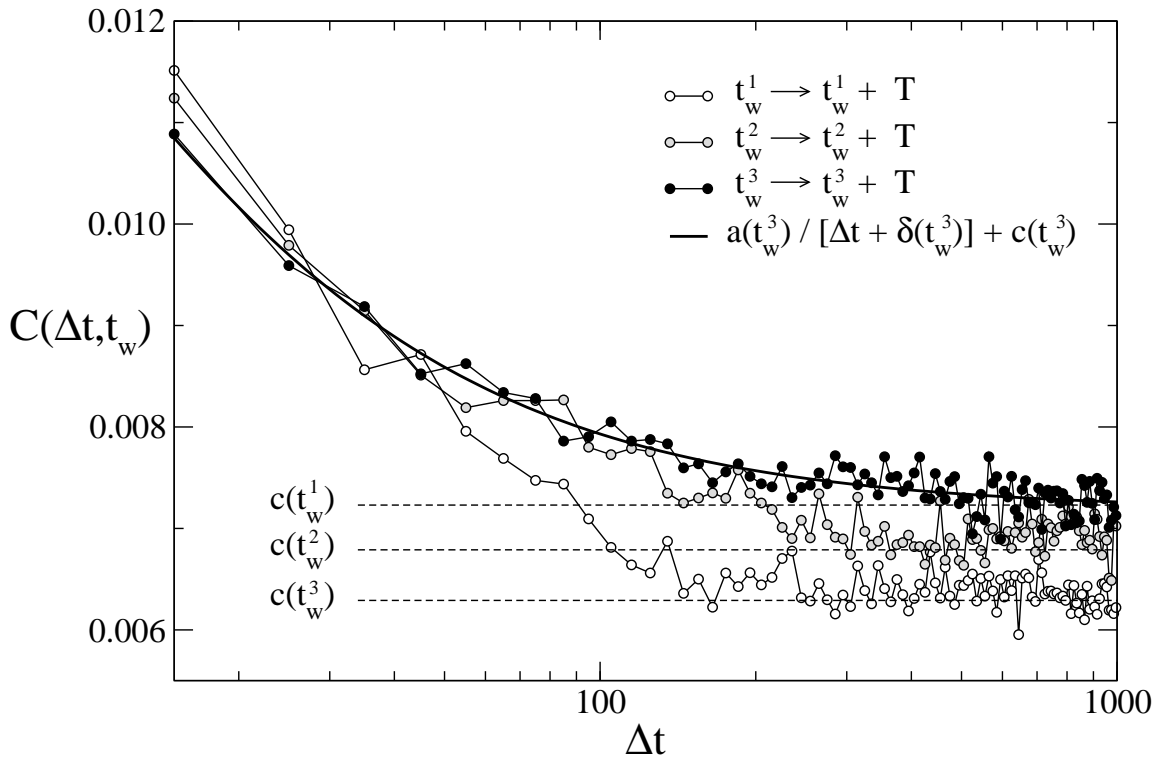}
\caption{
Tag-tag correlation functions and non-stationarity. The tag-tag correlation function $C(\Delta t, t_w)$
is computed over three consecutive and equally long ($T=30000$ tags each) subsets of the \textit{blog}
dataset, starting respectively at positions $t_w^1 = 10000$, $t_w^2 = 40000$ and $t_w^3=70000$ within
the collected sequence. Short-range correlations are clearly visible, slowly decaying towards
a long-range plateau value. The non-stationary character of correlations is visible both at short range,
where the value of the correlation function decays with $t_w$, and at long range,
where the asymptotic correlation increases with $t_w$. The long-range correlations (dashed lines) can be
estimated as the natural correlation present in a random sequence containing a finite number of tags:
on using the appropriate ranked distribution of tag frequencies within each window (see text)
the values $c(t_w^1)$, $c(t_w^2)$ and $c(t_w^3)$ can be computed, matching the measured plateau
of the correlation functions. The thick line is a fit to the fat-tailed memory kernel described in the text.
\label{fig2c}
}
\end{figure*}

We tested this hypothesis
with real data extracted from \textit{del.icio.us}: Fig.~\ref{fig2c}
shows the temporal auto-correlation function for the sequence of tags
co-occurring with \textit{blog}.
Such a sequence is constructed by consecutively appending the tags
associated with each post, respecting the temporal order of posts.
Correlations are computed inside
three consecutive windows of length $T$, starting at different times $t_w$,
\[
C(\Delta t, t_w) = \frac{1}{T - \Delta t}
\sum_{t=t_w+1}^{t=t_w+T-\Delta t} \,
\delta ( \mbox{tag}(t+\Delta t), \mbox{tag}(t) ) \, ,
\]
where $\delta ( \mbox{tag}(t+\Delta t), \mbox{tag}(t) )$ is the 
usual Kronecker delta function, taking the value $1$ when the same tag occurs
at times $t$ and $t+\Delta t$. From Fig.~\ref{fig2c} it is apparent
that the correlation function is non-stationary over time.
Moreover, for each value of the initial time $t_w$ a power-law behavior is observed:
$C(\Delta t, t_w) = a(t_w) / (\Delta t + \delta(t_w)) + c(t_w)$, where $a(t_w)$
is a time-dependent normalization factor and $\delta(t_w)$
is a phenomenological time scale, slowly increasing with the``age'' $t_w$
of the system.
$c(t_w)$ is the correlation that one would expect in a random sequence of tags
distributed according to the frequency-rank distribution $P_{T,t_w}(R)$
pertaining to the relevant data window.
Denoting by $R=R_{\mbox{max}}(T,t_w)$ the number of distinct tags occurring
in the window $[ t_w, t_w+T ]$,
we have $c(t_w) = \sum_{R=1}^{R=R_{\mbox{max}}(T,t_w)} \, P_{T,t_w}^{2}(R)$.

Our modification of the Yule-Simon's model thus consists in weighting
the probability of choosing an existing word (tag) according to a power-law kernel.
This hypothesis about the functional form of the memory kernel is also
supported by findings in Cognitive Psychology~\cite{b18}, where power laws
of latency and frequency have been shown to model human memory.

Summarizing, our model of users' behavior can be stated as follows:
the process by which users of a collaborative tagging system
associate tags to resources can be regarded as the construction of a ``text'',
built one step at a time by adding ``words'' (i.e. tags) to a text initially
comprised of $n_0$ words. This process is meant to model the behavior
of an effective average user in the context identified by a specific tag.
At a generic (discrete) time step $t$, a brand new word may be invented
with probability $p$ and appended to the text, while with probability $1-p$
one word is copied from the existing text, going back in time by $x$ steps
with a probability $Q_t(x)$ that decays as a power law, $Q_t(x) = a(t) / (x+\tau)$.
$a(t)$ is a normalization factor and $\tau$ is a characteristic time scale
over which recently added words have comparable probabilities.
Fig.~\ref{fig2b} shows the excellent agreement between the experimental data
and the numerical predictions of our Yule-Simon's model with long-term memory.
Our model, unsurprisingly, also reproduces the temporal correlation
behavior observed in real data (see \cite{nostro1} for a discussion of this point).

The interpretation of $\tau$ (similar to that of the $\delta$ parameter
introduced above for tag-tag correlations) 
is related to the number of equivalent top-ranked tags perceived by users
as semantically independent (see section \ref{highlowrank}).
In our model, in fact, the average user is exposed to a few roughly equivalent
top-ranked tags and this is translated mathematically into a low-rank cutoff
of the power law, i.e. the observed low-rank flattening.

Fitting the parameters of the model, in order to match its predictions (obtained by
computer simulation) against the experimental data, we obtain an excellent agreement
for all the frequency-rank curves we measured, as shown in Fig.~\ref{fig2b}.
This is a clear indication that the tagging behavior embodied in our simple model
captures some key features of the tagging activity.
The parameter $\tau$ controls the number of top-ranked tags which are allowed to co-occur
with comparable frequencies, so that it can be interpreted as a measure of the ``semantic breadth''
of a tag.  This picture is consistent with the fact that the fitted value of $\tau$ obtained
for \textit{blog} (a rather generic tag) is larger than the one needed
for \textit{ajax} (a pretty specific one).
Additional information on the role of $\tau$ as well as that of $p$
in the framework of our model are reported in \cite{nostro2}.

\begin{figure}[htb]
\includegraphics[keepaspectratio,width=\columnwidth]{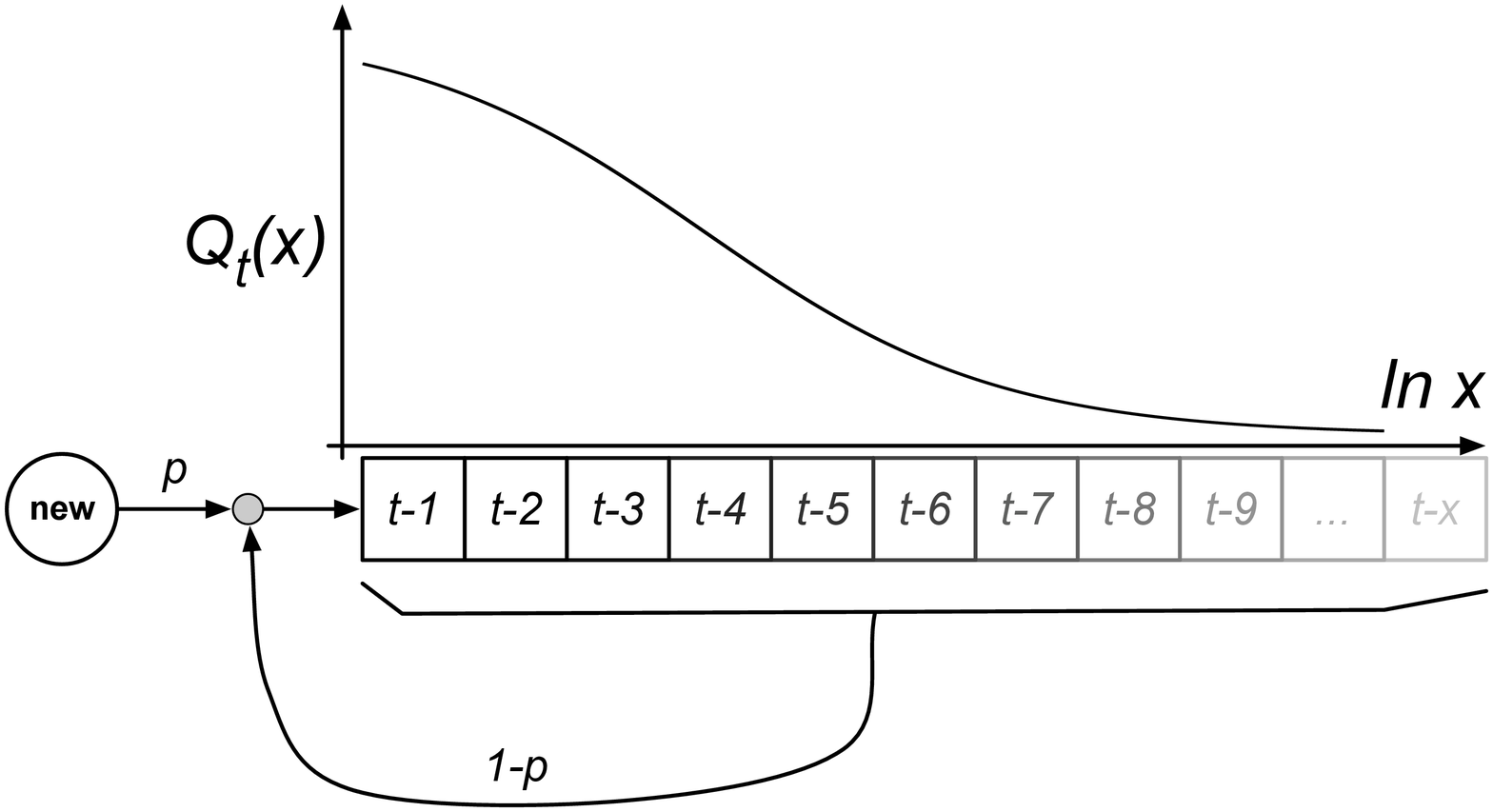}
\caption{
A Yule-Simon's process with long-term memory.
A synthetic stream of tags is generated by iterating the following step:
with probability $p$ a new tag is created and appended to the stream,
while with probability $1-p$ a tag is copied from the past of the stream
and appended to it. The probability of selecting a tag located $x$
steps into the past is given by the long-range memory kernel $Q_t(x)$,
which provides a fat-tailed access to the past of the stream.
\label{fig3}
}
\end{figure}

\newpage
\section{Co-occurrence between high-rank and low-rank tags}
\label{highlowrank}
\begin{figure*}
\vspace*{-0.5cm}
\hspace*{-0.6cm}
\includegraphics[width=1.05\textwidth,keepaspectratio,angle=0]{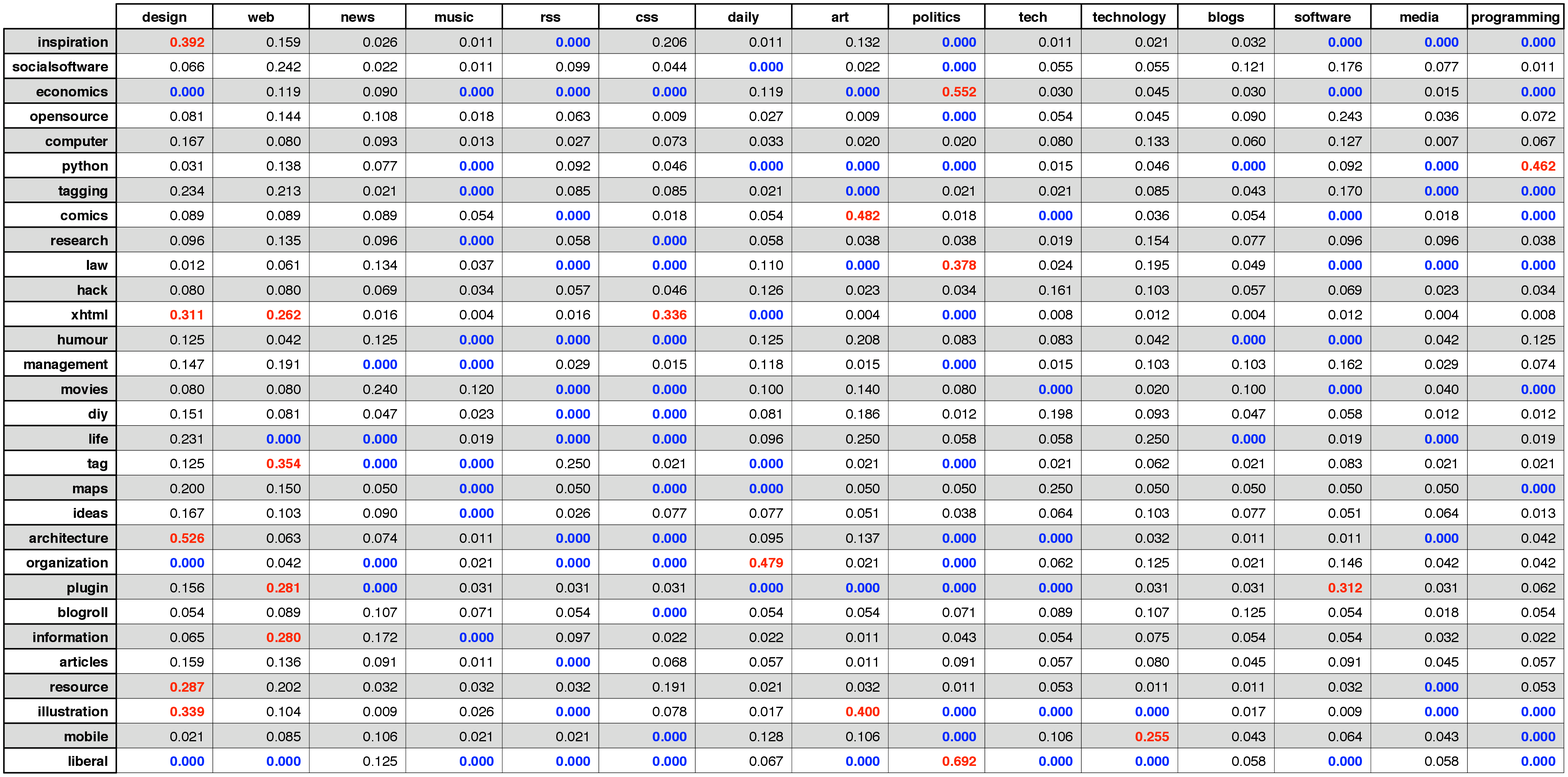}
\vspace*{-1.6cm}
\caption{
Co-occurrence table: columns correspond to the $15$ top-ranked tags co-occurring
with \textit{blog}, in descending order of frequency from left to right.
Rows correspond to $30$ low-frequency tags co-occurrinng with \textit{blog}
(frequencies ranking between $100$th and $200$th).
Each row is a normalized co-occurrence histogram
representing a ``categorization'' of the corresponding tag
in terms of the top-ranked tags.
Numbers in red (bold face) denote co-occurrence probabilities in excess of 25\%.
Zeros (no co-occurrence) are marked in blue (bold face).
\label{fig4}
}
\end{figure*}

\begin{figure*}
\vspace*{0.5cm}
\includegraphics[width=0.9\textwidth,keepaspectratio]{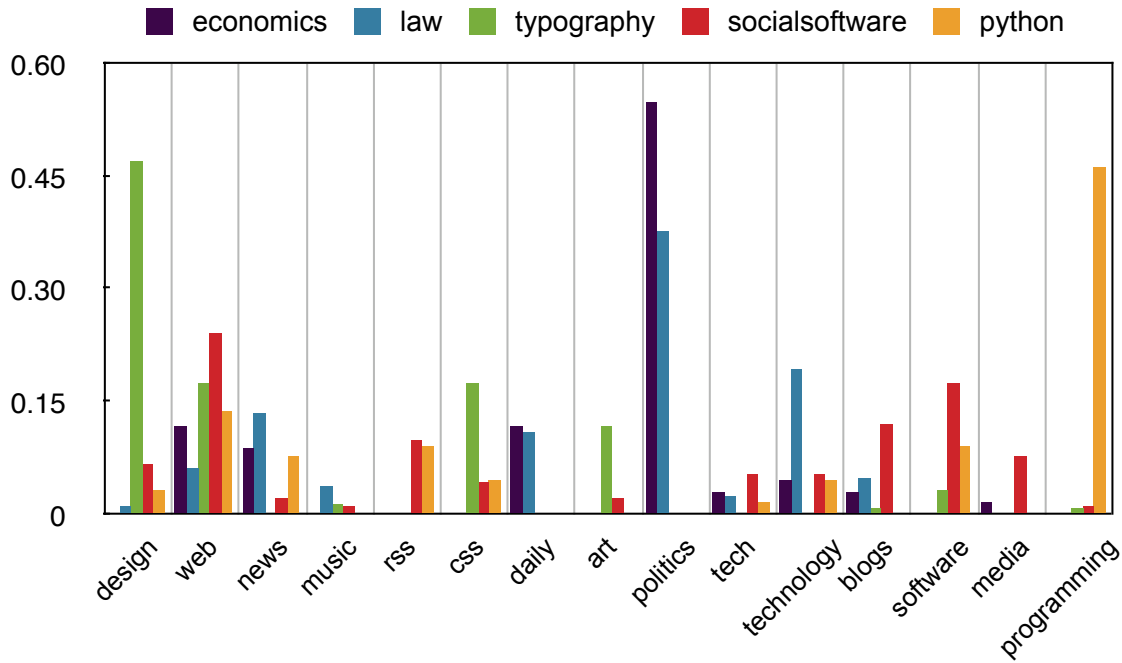}
\vspace*{-0.5cm}
\caption{
Co-occurrence patterns for $5$ of the low-frequency (high-rank) tags of Fig.~\ref{fig4}
(see legend at the top). 
The colored bars display the ``fingerprint'' of the selected tags in terms
of their co-occurrence with the $15$ top-ranked tags (the same ones reported
in the top row of Fig.~\ref{fig4}).
\label{fig5}
}
\end{figure*}

Fig.~\ref{fig4} shows a table where the occurrence of $30$
high-rank (low-frequency) tags is related to the occurrence
of the $15$ lowest-rank (highest-frequency) tags. All the tags under study
are co-occurring with the tag \textit{blog} and the dataset used for the analysis
is the same as the one used in Fig.~\ref{fig2a}.
The co-occurrence analysis is performed as follows: given a high-rank tag $X$,
all resources tagged with $X$ (within the above dataset) are selected,
and the co-occurrence frequencies of $X$ with each of the $15$ top-ranked
(most frequent) tags are recorded. Thus, each row of the table associates a tag $X$
with the corresponding (normalized) co-occurrence histogram.
This provides a statistical characterization of tag $X$
in terms of the top-ranked tags, regarded as a natural basis for categorization
(or semantic ``grounding'').
Fig.~\ref{fig5} graphically illustrates such a ``tag fingerprint''
for $5$ high-rank tags, arbitrarily chosen.
This analysis is aimed at probing the existence of non-trivial co-occurrence
relationships that might be ascribed to semantics and -- possibly -- to the emergence
of a self-organized hierarchy of tags. As shown by the bold numbers in Fig.~\ref{fig4},
as well as by the graph in Fig.~\ref{fig5}, high-frequency (low-rank) tags do not trivially co-occur
with most of the low-frequency (high-rank) tags --- on the contrary, the co-occurrence profile
of the latter is peaked in correspondence of specific, semantically related tags
(\textit{economics} and \textit{law} with \textit{politics}, for example, see Fig.~\ref{fig5}).
Moreover, several low-frequency (high-rank) tags never co-occur with some of the highest-frequency
(low-rank) tags,
as shown by the several zeros
in Fig.~\ref{fig4}. This suggests that high-frequency tags partition -- or ``categorize'' --
the resources marked by tags of lower frequency. Given that our definitions of ``high-rank''
and ``low-rank'' are somehow arbitrary, and given the self-similar character of tag
association we observed (Fig.~\ref{fig2b}), we expect our observations
to be representative of a general and complex semiotic structure underlying folksonomies.

\section{Conclusions}
Uncovering the mechanisms governing the emergence of shared
categorizations or vocabularies in absence of global coordination
is a key problem with significant scientific and technological potential.
Collaborative tagging provides a precious opportunity
to both analyze the emergence of shared conventions
and inspire the design of large (human or artificial) agent systems.
Here we report a statistical analysis of tagging activity in a popular social bookmarking system,
and introduce a simple stochastic model of user behavior which is able to reproduce
the measured co-occurrence properties to a surprisingly level of accuracy.
Our results suggest that users of collaborative tagging systems share
universal behaviors which, despite the intricacies of personal categorization,
tagging procedures and user interactions, appear to follow simple activity patterns.
In addition to the findings reported and discussed in this paper,
our approach constitutes a starting point upon which studies of greater complexity
can be based, with the final goal of understanding, predicting and controlling
the Semiotic Dynamics of online social systems.

\begin{acknowledgments}
The authors wish to thank A.~Baronchelli, A.~Baldassarri, V.~Servedio and L.~Steels for many interesting discussions and suggestions. This research has been partly supported by the ECAgents project funded by the Future and Emerging Technologies program (IST-FET) of the European Commission under the EU RD contract IST-1940. The information provided is the sole responsibility of the authors and does not reflect the Commission's opinion. The Commission is not responsible for any use that may be made of data appearing in this publication.
\end{acknowledgments}

\end{document}